\pdfoutput=1
\documentclass[%
reprint,
superscriptaddress, 
footinbib,
nobibnotes, 
amsmath,amssymb,
aps,
prl,
floatfix,
]{revtex4-2}

\usepackage{graphicx}
\usepackage{bm}
\usepackage{mathtools}
\usepackage{amsthm}
\usepackage{physics2,derivative}
\usephysicsmodule{ab,ab.braket}
\usepackage{xcolor}
\usepackage{natbib}
\usepackage[resetlabels]{multibib}
\newcites{SM}{References}

\makeatletter
\def\frontmatter@maketitle{%
  \@author@finish
  \title@column\titleblock@produce
  \suppressfloats[t]%
  \let\abstract\@undefined\let\endabstract\@undefined
  \titlepage@sw{%
   \vfil
   \clearpage
  }{}%
  \onecolumn@grid@setup
  \def\set@footnotewidth{\set@footnotewidth@one}%
}%
\makeatother

\renewcommand\paragraph[1]{%
\par\emph{#1.---}\kern2pt\relax\ignorespaces}

\theoremstyle{definition}
\newtheorem{theorem}{Theorem}

\newtheorem{corollary}[theorem]{Corollary}

\DeclareMathOperator{\Tr}{\mathrm{Tr}}
\DeclareMathOperator{\dist}{\mathrm{dist}}
\newcommand{\mc}{\mathrm{mc}}
\newcommand{\lEq}{\mathrm{leq}}
\newcommand{\energ}{U_0}
\newcommand{\vn}{S^{\mathrm{vN}}}

\usepackage[%
setpagesize=false,%
bookmarksnumbered=true,%
bookmarksopen=true,%
colorlinks=true,%
linkcolor=magenta,%
citecolor=blue,%
]{hyperref}
\hypersetup{pdfauthor={Akihiro Hokkyo and Masahito Ueda},pdftitle={Universal Upper Bound on Ergotropy and No-Go Theorem by the Eigenstate Thermalization Hypothesis}}

\begin{document}
\title{Universal Upper Bound on Ergotropy and No-Go Theorem by the Eigenstate Thermalization Hypothesis
}%

\author{Akihiro Hokkyo}
 \email{hokkyo@cat.phys.s.u-tokyo.ac.jp}
 \affiliation{Department of Physics, University of Tokyo, 7-3-1 Hongo, Bunkyo-ku, Tokyo, 113-8654, Japan}%
\author{Masahito Ueda}%
\affiliation{%
Department of Physics, University of Tokyo, 7-3-1 Hongo, Bunkyo-ku, Tokyo, 113-8654, Japan}
\affiliation{
 Institute for Physics of Intelligence, University of Tokyo, 7-3-1 Hongo, Bunkyo-ku, Tokyo, 113-0033, Japan}
 \affiliation{
 RIKEN Center for Emergent Matter Science (CEMS), Wako, 351-0198, Japan
}%

\begin{abstract}
  We show that the maximum extractable work (ergotropy) from a quantum many-body system is constrained 
  by \emph{local athermality} of an initial state and \emph{local entropy decrease} brought about by quantum operations. 
  The obtained universal upper bound on ergotropy implies that the eigenstate thermalization hypothesis prohibits work extraction from energy eigenstates by means of finite-time unitary operations. 
  This no-go property implies that Planck's principle, a form of the second law of thermodynamics, holds even for pure quantum states. 
  Our result bridges two independently studied concepts of quantum thermodynamics, the second law and thermalization, 
  via \textit{intrasystem} correlations in many-body systems as a resource for work extraction.
\end{abstract}

\maketitle
\paragraph{Introduction}
At the heart of quantum thermodynamics lies the problem of how thermodynamics emerges from microscopic dynamics. 
Recent advances in quantum control have enabled experimental exploration of this problem~\cite{binderThermodynamicsQuantumRegime2018}.
Highly controllable quantum systems with nearly complete isolation from environments, 
such as ultracold atomic gases and trapped ions, 
offer an ideal platform for experimental tests on the foundations of quantum thermodynamics and statistical mechanics. 
See Refs.~\cite{langenUltracoldAtomsOut2015,uedaQuantumEquilibrationThermalization2020} for reviews.

Investigation about whether isolated quantum systems thermalize dates back to von Neumann~\cite{neumann_beweis_1929}. 
Recent experiments have demonstrated thermalization in well-isolated quantum systems~\cite{kaufmanQuantumThermalizationEntanglement2016}. 
Theoretically, the eigenstate thermalization hypothesis (ETH)~\cite{deutschQuantumStatisticalMechanics1991,srednickiChaosQuantumThermalization1994,rigolThermalizationItsMechanism2008} 
has been proposed as a mechanism for thermalization in isolated quantum systems. 
The ETH states that energy eigenstates are thermal \textit{per se} through the lens of observables.
The ETH has been numerically verified for several different models~\cite{rigolThermalizationItsMechanism2008,kimTestingWhetherAll2014}.

Another central issue in thermodynamics is how much work can be extracted from a given system.
Planck's principle, which is a form of the second law of thermodynamics, 
expresses a no-go property on work extraction.
\emph{Passivity}~\cite{puszPassiveStatesKMS1978,lenardThermodynamicalProofGibbs1978,skrzypczykPassivityCompletePassivity2015}, 
which states that energy cannot be decreased by any unitary operation, 
was proposed as the quantum counterpart of Planck's principle. 
It is known~\cite{puszPassiveStatesKMS1978,lenardThermodynamicalProofGibbs1978} that 
the Gibbs state is passive, 
while pure states other than the ground state are not. 
Since the ETH states that energy eigenstates
are locally in thermal equilibrium, 
it is natural to ask how the ETH, which can be verified through observables, can be connected with the notion of thermal equilibrium in the sense of passivity.
A crucial observation here is that 
the original idea of passivity assumed that 
we can perform any unitary operation on the system, 
which requires unrealistic Hamiltonians involving nonlocal and $O(N)$-body interactions for $N\ (\gg1)$-particle systems. 
Passivity in many-body systems under realistic constraints on operations
and its connection with 
thermal equilibrium of observables
remain to be clarified. 
A closely related yet not fully explored subject concerns information thermodynamics, 
according to which feedback control allows one to extract an extra free energy beyond the conventional second law from a system in contact with a heat bath~\cite{sagawaSecondLawThermodynamics2008,sagawaMinimalEnergyCost2009,sagawaErratumMinimalEnergy2011}. 
Whether or not a similar work extraction can be made for a quantum system
that is isolated from heat baths and dissipation
deserves further study 
for a deeper understanding of the connection between information and quantum thermodynamics. 

The previous studies~\cite{kanekoWorkExtractionSingle2019,babaWorkExtractabilityEnergy2023} suggest that 
the ETH hinders work extraction from energy eigenstates by realistic unitary operations. 
However, a quantitative estimate of extractable work for general systems remains unexplored. 
Deriving a bound on the amount of extractable work from a given system is of fundamental importance in quantum thermodynamics, 
which is also important from an engineering point of view.
The primary purpose of this Letter is 
to derive a universal upper bound on the maximum work (ergotropy) 
that can be extracted from many-body systems with short-range interactions through general operations including feedback control. 
As a corollary, 
we show that the second law of thermodynamics holds even for pure states in thermal equilibrium with respect to observables.

\paragraph{Ergotropy and passivity}
Let us first introduce the extractable work from a quantum state by means of quantum operations, 
which is known as ergotropy~\cite{allahverdyanMaximalWorkExtraction2004}. 
For a Hamiltonian $H$ and an initial state $\rho$, 
the ergotropy by a quantum channel $f$ 
and that by a class of channels $\mathcal{F}$ are defined as
\begin{align}
  W_{f,H}(\rho)&\coloneqq\braket<H>_{\rho}-\braket<H>_{f(\rho)}, \\
  W_{\mathcal{F},H}(\rho)&\coloneqq \sup_{f\in\mathcal{F}}W_{f,H}(\rho),
\end{align}
where $\braket<H>_{\rho}\coloneqq \Tr(H\rho)$. 
Passivity was originally defined as the non-positivity of the ergotropy
by $\mathcal{F}$ that is constituted of all unitary maps~\cite{puszPassiveStatesKMS1978,lenardThermodynamicalProofGibbs1978}. 
Several works~\cite{freyStrongLocalPassivity2014,
brownPassivityPracticalWork2016,mukherjeePresenceQuantumCorrelations2016,
onuma-kaluWorkExtractionUsing2018,
alimuddinBoundErgotropicGap2019,alhambraFundamentalLimitationsLocal2019,
senLocalPassivityEntanglement2021,
mitsuhashiCharacterizingSymmetryProtectedThermal2022,biswasExtractionErgotropyFree2022,
salviaOptimalLocalWork2023,safranekWorkExtractionUnknown2023,koshiharaQuantumErgotropyQuantum2023} 
have investigated the generalization of passivity to $\mathcal{F}$ 
that is different from the set of all unitary maps.
In the following, we first derive the ergotropy bounds for general operations in many-body systems 
and then apply the bounds for operations subject to realistic constraints. 

\paragraph{Models}
We consider quantum spin systems on a $D$-dimensional hypercubic lattice with fixed (periodic or open) boundary conditions; 
a similar discussion can be made for spin systems on a more general lattice or a graph. 
We denote a set of lattice sites as $\Lambda=\{1,2,\ldots, L\}^D$ 
and the system size as $V=L^D$. 
The distance between sites $i$ and $j$ is denoted by $r_{ij}$, 
which is defined as the Euclidean distance subject to the boundary conditions. 
For each site $i\in\Lambda$, we have a $d=(2\mathcal{S}+1)$-dimensional state space $\mathcal{H}_i\cong\mathbb{C}^d$, 
where $\mathcal{S}$ is the spin quantum number. 
For a subsystem $A\subset\Lambda$, 
the reduced state of a state $\rho$ is denoted as 
$\rho_A\coloneqq \Tr_{\Lambda\setminus A}\rho$. 

Next, we introduce a Hamiltonian of our system. 
Considering only two-body interactions, the Hamiltonian is expressed as 
$H=\sum_{i\in\Lambda}h_i+\sum_{i\neq j\in\Lambda}U_{ij}$. 
Here, $h_i$ is the on-site Hamiltonian acting on $\mathcal{H}_i$, 
and $U_{ij}$ is a two-body interaction with $U_{ij}=U_{ji}$ acting on $\mathcal{H}_i\otimes\mathcal{H}_j$.
We assume that $h_i\eqqcolon U_{ii}$ is bounded and $U_{ij}$ decays sufficiently fast as
\begin{equation}
\|U_{ij}\|\leq \energ(1+r_{ij})^{-(D+\delta)}\ (\forall i,j\in\Lambda). \label{short-range}
\end{equation}
Here, $\energ$ and $\delta$ are positive constants independent of $V$. 
This property guarantees the additivity of energy. 
For simplicity of notation, we restrict the model to spin systems. 
However, as shown later, our results are applicable to bosonic and fermionic systems on a lattice with slight modifications. 

\paragraph{Universal upper bound on ergotropy}
We divide the lattice into small hypercubes: $\Lambda=\bigsqcup_{A\in\mathcal{A}}A$~\footnote{
  For the following discussion, 
  it is sufficient if $|\Lambda\setminus\bigsqcup_{A\in\mathcal{A}}A|=o(V)$ without $\mathcal{A}$ being an exact partition. 
  See Supplemental Material~\cite{suppl}. 
}, 
and assume that the linear dimension $l$ of each $A$ depends on $V$ and diverges as $V\to\infty$. 
The Hamiltonian $H_A$ on each subsystem $A$ and the residual interaction $U^R_{\mathcal{A}}$ are defined as 
\begin{align}
  H_A&\coloneqq \sum_{i\in A}h_i+\sum_{i,j\in A,\ i\neq j}U_{ij},\\
  U^R_{\mathcal{A}}&\coloneqq H-\sum_{A\in\mathcal{A}}H_A=\sum_{A\in\mathcal{A}}\sum_{i\in A,j\notin A}U_{ij}.
\end{align}
It follows from Eq.~\eqref{short-range} and the assumption $l\to\infty$ 
that $U^R_{\mathcal{A}}$ is subextensive in the thermodynamic limit~\cite{suppl}. 
In the following, 
an equality and an inequality of energy within the accuracy of $U_0\times o(V)$ are denoted by $\simeq$ and $\lesssim$, respectively. 

Let $E_{H_A}$ be the energy of the canonical state on $A$ defined as~\footnote{
  This is just a Legendre transformation of free energy. 
  See Supplemental Material~\cite{suppl} for details. 
} 
\begin{equation}
  E_{H_A} (S)\coloneqq \sup_{\beta>0}\left[\beta^{-1}\left(-\ln\Tr e^{-\beta H_A}+S\right)\right], 
\end{equation}
where $S\ (<|A|\ln d)$ is an arbitrary variable corresponding to entropy. 
We also introduce the temperature of $A$ as $\beta_A^{-1}(S)\coloneqq \pdv{}{S}E_{H_A} (S)$. 

We are now in a position to state the main result of this Letter: the universal upper bound on ergotropy. 
Let the energy of the initial state $\rho$ be $V\epsilon$ 
and consider the \textit{local (thermal) equilibrium ensemble} $\rho^\lEq$ 
that satisfies the following conditions: 
\begin{align}
  &V\epsilon\simeq\braket<H>_{\rho^\lEq}\simeq \sum_AE_{H_A} (\vn(\rho^\lEq_A))
  , \label{condition_ensemble}\\
  &(\beta_A^\lEq)^{-1}\coloneqq \beta_A^{-1}(\vn(\rho^\lEq_A))\leq\beta_0^{-1}\ (\forall A\in\mathcal{A}), \label{condition_subsystems}
\end{align}
where $\vn$ is the von Neumann entropy, and $\beta_0>0$ is a positive constant independent of $V$.
The first condition~\eqref{condition_ensemble} is satisfied for equilibrium statistical ensembles~\cite{ruelleStatisticalMechanicsRigorous1999,tasaki_local_2018} 
such as microcanonical and canonical ensembles 
and the product of local Gibbs states of each subsystems $A\in\mathcal{A}$ 
which, in general, have different temperatures $(\beta_A^{-1})_{A\in\mathcal{A}}$. 
The second condition~\eqref{condition_subsystems} indicates that the temperature of the subsystem is not too high. 
When the Hamiltonian is translationally invariant, 
this condition is valid 
unless $\epsilon$ is too large. 
A detailed analysis of these conditions is made in Supplemental Material~\cite{suppl}. 
\begin{theorem}\label{thm:universal_bound}
  We consider a general class of operations $\mathcal{F}$ including the identity map.  
  Under conditions~\eqref{condition_ensemble} and~\eqref{condition_subsystems}, 
  the ergotropy is bounded from above as 
  \begin{align}
    &W_{\mathcal{F},H}(\rho)\lesssim 
    \sum_A \ab[E_{H_A} (\vn(\rho^\lEq_A))-E_{H_A} (\vn(\rho_A))]\nonumber\\
    &\ \ \ +\sup_{f\in\mathcal{F}}\sum_A\ab[E_{H_A} (\vn(\rho_A))-E_{H_A}(\vn(f(\rho)_A))],\label{universal_bound}
    \end{align}
where both terms on the right-hand side (r.h.s.)~are non-negative. 
The first term is further bounded from above as
\begin{align}
  \sum_A &\ab[E_{H_A} (\vn(\rho^\lEq_A))-E_{H_A} (\vn(\rho_A))]\nonumber\\
  &\lesssim V\beta_0^{-1}(\ln d)\max_{A\in\mathcal{A}}\|\rho^\lEq_A-\rho_A\|_1\label{local_athermality}.
\end{align}
\end{theorem}
Theorem~\ref{thm:universal_bound} decomposes the upper bound on the work extractable from a many-body system into two terms having different physical meanings. 
The first term on the r.h.s.~of inequality~\eqref{universal_bound} characterizes the \emph{local athermality} of the initial state.
If we take $\rho^\lEq$ as a statistical ensemble, 
the r.h.s.~of inequality~\eqref{local_athermality}  
is an indicator of microscopic thermal equilibrium (MITE)~\cite{popescuEntanglementFoundationsStatistical2006,goldsteinSecondLawThermodynamics2013}  
and provides a measure of how local observables deviate from their thermal-equilibrium values. 
For general $\rho^\lEq$, this term represents the deviation of the initial state $\rho$ from local thermal equilibrium. 
The second term on the r.h.s.~of inequality~\eqref{universal_bound} gives the contribution from a decrease in entropy of the subsystem by means of quantum operations in $\mathcal{F}$. 
For pure initial states, 
it is necessary to break the correlation between the subsystem and the rest of the system 
by a quantum operation for the extraction of positive work.
In this sense, the second term may be interpreted as work gained from information encoded in intrasystem correlations, 
which is analogous to the results obtained in noninteracting systems~\cite{funoThermodynamicWorkGain2013,perarnau-llobetExtractableWorkCorrelations2015,salviaExtractingWorkCorrelated2022,touilErgotropyQuantumClassical2022}. 
Our result indicates a close relationship among ergotropy in a many-body system, thermalization of local observables, 
and information stored within the system that can be utilized by quantum operations, 
independent of specific models, initial states, and operations.

Before we prove the theorem, 
let us discuss an example for which the second term on the r.h.s.~of inequality~\eqref{universal_bound} gives the main contribution.
We consider a 1D Ising chain subject to the periodic boundary condition 
and write the Hamiltonian as $H=-\sum_{i=1}^Ls_i^zs_{i+1}^z-h\sum_{i=1}^Ls_i^z$
where $s^z_i=\ket|0>_i\bra<0|_i-\ket|1>_i\bra<1|_i$. 
We divide the whole system into subsystems, each of which has $l\ (\leq L/2)$ consecutive sites. 
We assume that the initial state $\ket|\Psi(\lambda)>$ is given by the product of the following long-range entangled states:
\begin{equation}
  \ket|\Psi(\lambda)>_i=\sqrt{1-\lambda}\ket|0>_i\ket|0>_{i+\frac{L}{2}}+\sqrt{\lambda}\ket|1>_{i}\ket|1>_{i+\frac{L}{2}}
\end{equation}
for $1\leq i\leq L/2$ and $0\leq\lambda\leq1/2$, where we assume $L$ is even. 
For $\lambda=1/2$, 
it is known that such long-range entangled states can be exact energy eigenstates of nonintegrable spin chains~\cite{udupaWeakUniversalityQuantum2023,ivanovVolumeentangledExactEigenstates2024,chibaExactThermalEigenstates2024}.
The subsystem entropy and energy density of $\ket|\Psi(\lambda)>$ are $lH_2(\lambda)$ 
and $-(1-1/l)(1-2\lambda)^2-h(1-2\lambda)$, respectively, 
where $H_2(\lambda):=-\lambda \ln \lambda-(1-\lambda)\ln(1-\lambda)$ is the binary entropy. 
As shown in Fig.~\ref{fig:thermal_pure}, 
this initial state is almost in local thermal equilibrium. 
Therefore, we need to decrease the entropy of the subsystem extensively to extract extensive work. 
We can achieve this extensive decrease in entropy as follows.
Consider a nonlocal CNOT gate 
that transforms the state as $\ket|\Psi(\lambda)>_i\mapsto(\sqrt{1-\lambda}\ket|0>_{i}+\sqrt{\lambda}\ket|1>_{i})\ket|0>_{i+\frac{L}{2}}$; 
then half of the subsystems reach the ground state. 
Moreover, all the entropies of the subsystems vanish. 
If we further perform unitary transformations on each site $i$, 
the entire system becomes the ground state, and the equality of Theorem~\ref{thm:universal_bound} is achieved. 
While local measurement and feedback can do the same job, 
measurement is not needed here in order to exploit the contribution of the second term on the r.h.s.~of \eqref{universal_bound}. 
\begin{figure}[tbp]
  \hspace*{-1cm}
  \includegraphics[width=\linewidth]{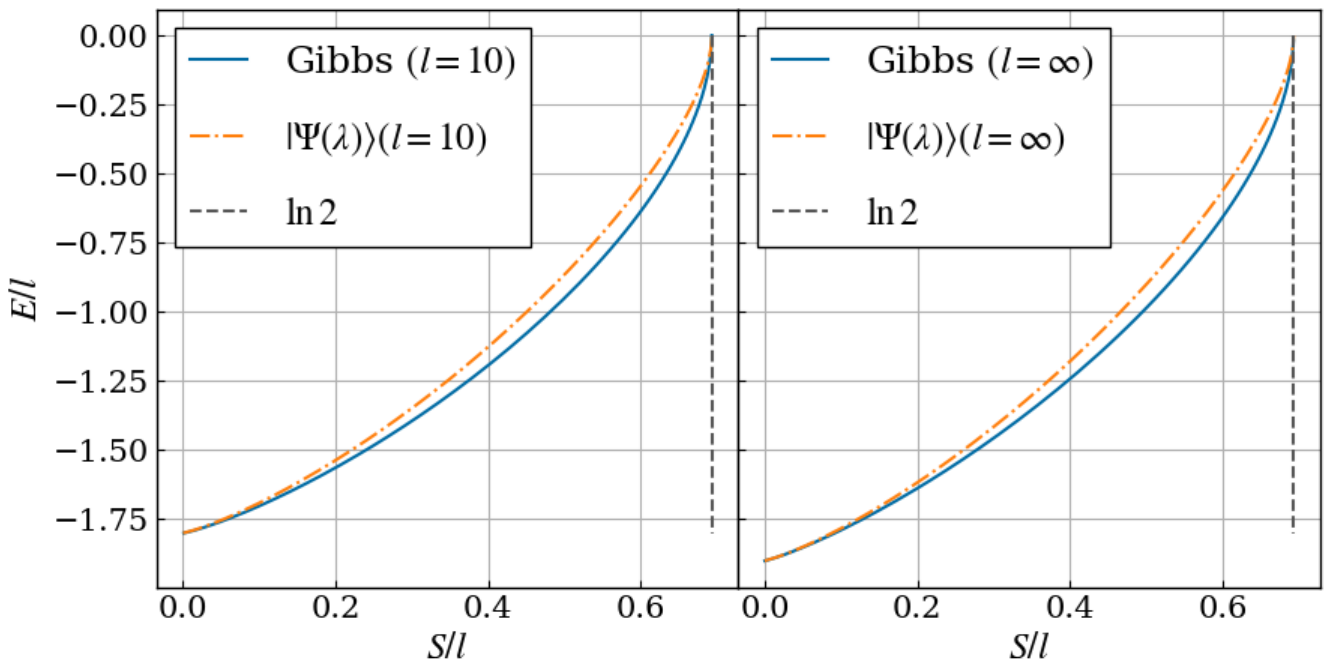}
  \caption{Energy density $E/l$ for $h=0.9$ versus entropy density $S/l$ of a subsystem of size $l=10$ and $l=\infty$ (thermodynamic limit). 
  The solid curves are obtained for a set of the Gibbs states of the subsystem with positive temperatures, 
  and the dash-dotted curves are obtained for $\{\ket|\Psi(\lambda)>\mid0\leq\lambda\leq1/2\}$. 
  We note that these values are independent of $L$ as long as $l\leq L/2$.
  The distance between the two curves shows the first term on the r.h.s.~of inequality~\eqref{universal_bound}, 
  which is much smaller than the energy density $E/l$ itself.
  }
  \label{fig:thermal_pure}
\end{figure}

Let us now prove Theorem~\ref{thm:universal_bound}. 
First, we show inequality~\eqref{universal_bound}. 
By condition~\eqref{condition_ensemble}, 
it is sufficient to show the following inequality:
\begin{equation}
  E_{H_A}(\vn(\sigma))
  \leq\braket<H_A>_\sigma \label{inequalities_maxent}
\end{equation}
for an arbitrary state $\sigma$.
This inequality can be proven from the property of the Gibbs state~\cite{suppl}, which is analogous to the maximum entropy principle~\cite{jaynesInformationTheoryStatistical1957}. 
We also find the non-negativity of the first term on the r.h.s.~of \eqref{universal_bound} by setting $\sigma=\rho$. 
The non-negativity of the second term is obvious since $\mathcal{F}$ includes the identity map. 

Next, we show inequality~\eqref{local_athermality}. 
Using the convexity of $E _{H_A}$ and the Fannes inequality~\cite{fannesContinuityPropertyEntropy1973,nielsenQuantumComputationQuantum2010}, 
the left-hand side (l.h.s.)~of inequality~\eqref{local_athermality} is bounded as 
\begin{align}
  &(\mbox{l.h.s.~of~\eqref{local_athermality}})
  \leq\sum_{A\in\mathcal{A}}{(\beta_A^\lEq)^{-1}}\left(\vn(\rho^\lEq_A)-\vn(\rho_A)\right)\nonumber\\
  \leq&\sum_{A\in\mathcal{A}}{(\beta_A^\lEq)^{-1}}\left[|A|(\ln d)\|\rho^\lEq_A-\rho_A\|_1+1/e\right]. \label{average_athermality}
\end{align}
Using condition~\eqref{condition_subsystems},  
we can further evaluate the r.h.s.~of~\eqref{average_athermality} as 
\begin{equation}
  (\mbox{r.h.s.~of~\eqref{average_athermality}})
  \lesssim V\beta_0^{-1}(\ln d)\max_{A\in\mathcal{A}}\|\rho^\lEq_A-\rho_A\|_1. \label{prf_1}
\end{equation}
This completes the proof of Theorem \ref{thm:universal_bound}. 
\paragraph{No-go theorem by the ETH}
When the contribution of entropy decrease is negligible, 
ergotropy is suppressed only by local athermality. 
This is the case for on-site unitary operations $\otimes_i\mathcal{U}_i$. 
More realistic maps are unitary evolutions by time-dependent Hamiltonians satisfying~\eqref{short-range}, 
which we shall refer to as local control. 
In fact, the small incremental entangling theorem~\cite{bravyiUpperBoundsEntangling2007,vanacoleyenEntanglementRatesArea2013} indicates that 
the rate of change in $S_A$ follows the area law~\cite{marienEntanglementRatesStability2016,gongEntanglementAreaLaws2017}, 
and hence it is at most $|\partial A|\sim l^{D-1}$. 
An analysis similar to what is made in the previous paragraph shows that 
the second term on the r.h.s.~of inequality~\eqref{universal_bound} for the operation time $T$ (per $\hbar$) is bounded from above by 
\begin{equation}
  \beta_1^{-1}\sup_{f\in\mathcal{F}}\sum_{A\in\mathcal{A}}|\vn(\rho_A)-\vn(f(\rho)_A)|\leq \beta_1^{-1}U_0To(V), 
\end{equation}
where we assume that the temperature of every subsystem is bounded from above by a $V$-independent positive temperature $\beta_1^{-1}$,
\begin{equation}
  \beta_A^{-1}(\vn(\rho_A))\leq \beta^{-1}_1\ (\forall A\in\mathcal{A}). \label{additional_assumption}
\end{equation}
Then, under the same conditions assumed in Theorem~\ref{thm:universal_bound}, we have the following corollary.
\begin{corollary}\label{cor:MITE_means_passive}
  Suppose that the initial state $\rho$ satisfies~\eqref{additional_assumption}.
  Let $\mathcal{F}$ be a class of local controls satisfying $U_0T=O(1)$. 
  Then, we have 
  \begin{equation}
    W_{\mathcal{F},H}(\rho)\lesssim
    V\beta_0^{-1}(\ln d)\max_{A\in\mathcal{A}}\|\rho^\lEq_A-\rho_A\|_1. 
  \end{equation}
  If $\rho$ is in $l$-local MITE~\cite{moriThermalizationPrethermalizationIsolated2018}, 
  then the r.h.s.~is subextensive 
  since the trace distance decays exponentially with increasing $V$.
  In particular, 
  if the ETH holds for all $l$-local observables, 
  it is impossible to extract the extensive work from energy eigenstates 
  by finite-time local control.
\end{corollary}
We refer to $\rho$ as \textit{thermodynamically passive} for $\mathcal{F}$ if $W_{\mathcal{F},H}(\rho)\lesssim0$. 
This corollary shows that,
even if the initial state is pure, 
it is thermodynamically passive for $\mathcal{F}$ as long as local observables are in thermal equilibrium. 
This is in contrast to the conventional passivity, 
and our result gives a stronger bound than that obtained in Refs.~\cite{puszPassiveStatesKMS1978,lenardThermodynamicalProofGibbs1978}. 
There are a few previous studies on work extraction from pure states and energy eigenstates in many-body systems~\cite{tasakiSecondLawThermodynamics2000,goldsteinSecondLawThermodynamics2013,kanekoWorkExtractionSingle2019,babaWorkExtractabilityEnergy2023}.
For a fixed unitary map, extensive work cannot be extracted from pure states involving a large number of energy eigenstates in the energy shell, if the initial time of operation is chosen at random~\cite{tasakiSecondLawThermodynamics2000,goldsteinSecondLawThermodynamics2013}. 
The authors of Ref.~\cite{kanekoWorkExtractionSingle2019} 
numerically explored the number of energy eigenstates from which extensive work can be extracted in a single quench operation
and also analytically showed that the fraction of such eigenstates decays exponentially for general systems. 
While these works consider thermodynamic passivity for a single unitary operation, 
our results can treat a general class of operations. 
Furthermore, in contrast to the analytical results of previous studies~\cite{tasakiSecondLawThermodynamics2000,goldsteinSecondLawThermodynamics2013,kanekoWorkExtractionSingle2019}, 
we consider fixed initial states and operations.
In a related study~\cite{babaWorkExtractabilityEnergy2023}, 
the number of work-extractable energy eigenstates by a state-by-state optimized operation was investigated. 
Corollary~\ref{cor:MITE_means_passive} analytically and independently of specific models justifies their numerical result, 
which states that work cannot be extracted by local control from those eigenstates that satisfy the ETH.

A couple of comments on this corollary are in order here.
Although the assumption in Corollary~\ref{cor:MITE_means_passive} is satisfied for an arbitrary finite time, 
it must be thermodynamically short. 
In fact, if thermalization occurs, 
it takes at least $U_0T\gg O(1)$~\cite{gongBoundsNonequilibriumQuantum2022} for the entropy of the subsystem to change extensively. 
If it takes an exponentially long time to reduce entropy as discussed in Ref.~\cite{babaWorkExtractabilityEnergy2023},  
then thermodynamic passivity can be shown for such timescales. 
The other comment is about complete passivity, 
which represents passivity for the direct product of any number of identically copied states~\cite{puszPassiveStatesKMS1978,lenardThermodynamicalProofGibbs1978,alickiEntanglementBoostExtractable2013,perarnau-llobetMostEnergeticPassive2015}. 
Unlike the original definition for a fixed finite system, 
we consider here the thermodynamic limit. 
Therefore, if copied systems interact without changing a spatial dimension $D$, 
no significant distinction arises between thermodynamic passivity and \textit{thermodynamic complete passivity}. 

\paragraph{Achievability of upper bound on ergotropy}
The condition for our inequality \eqref{universal_bound} to achieve the equality is
that $f(\rho)$ is locally in thermal equilibrium for an operation $f$ that achieves the supremum. 
This is due to the condition for achieving equality of the second inequality of \eqref{inequalities_maxent}.
In particular, if $H$ satisfies the (off-diagonal) ETH, 
since the system thermalizes after a long time,
the equality can always be achieved 
if we consider a class $\mathcal{F}$ 
such that we wait for a long time after the control of the Hamiltonian is completed.
For general operations, 
there is a gap between the ergotropy and the upper bound due to the nonequilibrium property of the final state.
If this could be evaluated from below, 
a \textit{lower} bound for ergotropy may also be obtained.

\paragraph{Applicability to fermionic and bosonic systems}
So far, we have restricted our discussion to spin systems.
However, the same result holds for fermionic systems and some bosonic systems.
For example, Theorem~\ref{thm:universal_bound} and Corollary~\ref{cor:MITE_means_passive} both hold for the following class of Hamiltonians, including the Fermi--Hubbard model: 
\begin{equation}
  H=\sum_{ij,\sigma}t_{ij}c_{i\sigma}^\dag c_{j\sigma}+\sum_{ij}U_{ij}n_in_j+\sum_{ij}U'_{ij}\bm{S}_i\cdot\bm{S}_j, 
\end{equation}
where $\sigma$ represents the spin of particles, $n_i=\sum_\sigma c_{i\sigma}^\dag c_{i\sigma}$ is the number operator and $\bm{S}_i$ is the second-quantized spin operator.
Here, $t,U,$ and $U'$ are assumed to decay similarly as in~\eqref{short-range}. 

For bosonic systems, we need to take into account the lack of limitation on the local number of particles. 
Our results can be generalized to the Bose--Hubbard model using recent work on information propagation in interacting boson systems~\cite{kuwaharaEffectiveLightCone2024}.
A detailed discussion on these generalizations is made in Supplemental Material~\cite{suppl}.

\paragraph{Conclusion and outlooks}
In this Letter, 
we have shown that work that can be extracted from a system is bounded from above by local athermality and local entropy decrease. 
In many-body systems, 
our results demonstrate that in addition to nonequilibrium properties, the \emph{information} stored in the system can also serve as a resource for work extraction.
Moreover, 
through the short-time evolution by a time-dependent Hamiltonian with short-range interactions, 
no extensive work can be extracted from energy eigenstates if the ETH holds. 
This result is consistent with previous numerical studies~\cite{kanekoWorkExtractionSingle2019,babaWorkExtractabilityEnergy2023}
on specific systems where the ETH is \emph{believed} to hold. 
In contrast, this Letter rigorously shows that the ETH prohibits work extraction for general interacting systems. 

While we briefly discussed lower bounds for ergotropy in the previous section, 
it is highly nontrivial to obtain a concrete expression for the lower bound except when the equality is achieved.
Such a study potentially leads to a generalization of information thermodynamics in many-body physics.

Another interesting extension is the no-go theorem in a long-time regime.
Whether our derivation of the second law can be extended to the long-time regime deserves further study.
Under appropriate additional assumptions, Corollary~\ref{cor:MITE_means_passive} is expected to hold in the long-time regime 
because the system relaxes to local equilibrium in a very short time~\cite{reimannTypicalFastThermalization2016} 
and hydrodynamics is applicable when the initial state is in local equilibrium~\cite{hayataRelativisticHydrodynamicsQuantum2015}. 
Such behavior of many-body systems in the short-time regime is also experimentally accessible.~\cite{leObservationHydrodynamizationLocal2023}.

While we consider the ETH for subsystems, the connection to a more general form of the ETH,  
such as one characterized by few-body observables~\cite{moriThermalizationPrethermalizationIsolated2018}, 
remains elusive. 
Clarifying the relationship between the observables we measure and the class of operations that satisfy the second law 
is not only of fundamental importance but also of practical significance, 
as it may lead to the resolution of the problem of how to extract work from quantum many-body systems with high efficiency 
\textit{beyond} the limitations of macroscopic thermodynamics.

\paragraph{Acknowledgement}
We are very grateful to Koki Shiraishi for the helpful discussions on the entangling power. 
We also thank Masaya Nakagawa and Shoki Sugimoto for their valuable comments on our draft.
This work was supported by KAKENHI Grant No. JP22H01152 from the Japan Society for the Promotion of Science and the CREST program ``Quantum Frontiers" (Grant No. JPMJCR23I1) from the Japan Science and Technology Agency. 
A.H. is supported by Forefront Physics and Mathematics Program to Drive Transformation (FoPM), a World-Leading Innovative Graduate Study (WINGS) Program, the University of Tokyo.

\bibliographystyle{apsrev4-2}
\bibliography{ergotropy,suppl}

\clearpage\clearpage
\makeatletter
   	\c@secnumdepth=4
\makeatother

\setcounter{equation}{0}
\setcounter{figure}{0}
\setcounter{section}{0}
\setcounter{table}{0}
\renewcommand{\theequation}{S\arabic{equation}}
\renewcommand{\thefigure}{S\arabic{figure}}
\renewcommand{\theHequation}{\theequation}
\renewcommand{\theHfigure}{\thefigure}

\renewcommand{\bibnumfmt}[1]{[S#1]}
\renewcommand{\citenumfont}[1]{S#1}

\renewcommand{\thepage}{S\arabic{page}}
\setcounter{page}{1} 

\title{
  Supplemental Material: \protect\\
  Universal Upper Bound on Ergotropy and No-Go Theorem by the Eigenstate Thermalization Hypothesis
}
\date{\today}
\maketitle
\onecolumngrid

We present a few technical details omitted in the main text. 
These are all well-known results in the theory of rigorous statistical mechanics or quantum many-body systems, 
but for the sake of self-containedness we provide rigorous claims and give some proofs thereof.
In Sec.~\ref{sec:suppl_short-range}, the results derived from the short-range property of the interaction are presented, 
and in Sec.~\ref{sec:suppl_canonical}, we give the definitions and the formulae of the statistical-mechanical quantities.
In Sec.~\ref{sec:suppl_FBsystem}, we discuss the generalization of our results in spin systems to fermionic and bosonic systems.

\section{Consequences from short-range interactions}\label{sec:suppl_short-range}
\subsection{Residual interaction}
We show that the residual interaction is subextensive
when the interaction decays as
\begin{equation}
  \|U_{ij}\|\leq U_0(1+r_{ij})^{-(D+\delta)}\ (\forall i,j\in\Lambda). \label{suppl:short-range}
\end{equation}
Since the residual interaction is decomposed as
\begin{equation}
\|U^R_{\mathcal{A}}\|\leq\sum_{A\in\mathcal{A}}\sum_{i\in A,j\in\Lambda\setminus A}\|U_{ij}\|\eqqcolon \sum_{A\in\mathcal{A}}U_A, \label{suppl:residual}
\end{equation}
it is sufficient to show $\sup_{A\in\mathcal{A}}U_A=o(|A|)=o(l^D)$. 

We can take $\kappa_D>0$ (independent of $V$) satisfying
\begin{equation}
  \sum_{\substack{j\in\Lambda\\r-1<r_{ij}\leq r}}1\leq \kappa_Dr^{D-1}\ (\forall i\in\Lambda,r\in\mathbb{Z}_{>0}),
\end{equation}
because we have 
\begin{equation}
  C_D(r-\sqrt{D})^D\leq
  2^D\sum_{\substack{j\in(\mathbb{Z}_{>0})^D\\ \|j\|_2\leq r}}1\leq
  \sum_{\substack{j\in\mathbb{Z}^D\\\|j\|_2\leq r}}1\leq 
  2^D\sum_{\substack{j\in(\mathbb{Z}_{\geq0})^D\\\|j\|_2\leq r}}1\leq
  C_D(r+\sqrt{D})^D, 
\end{equation}
where $\|\bullet\|_2$ is the Euclidean norm and $C_D$ is the volume of the unit ball with dimension $D$. 
Also, we have $\kappa'_D>0$ (independent of $V,l$) satisfying
\begin{equation}
  \sum_{\substack{i\in A\\ \dist(i,\Lambda\setminus A)\leq r}}1\leq\kappa'_Dl^{D-1}r\ (\forall r\in\mathbb{Z}_{>0}), \label{suppl:bulk_property}
\end{equation}
where $\dist(i,B)\coloneqq \inf_{j\in B}r_{ij}$. 
Inequality~\eqref{suppl:bulk_property} can be shown from the following inequalities: 
\begin{align}
  \sum_{\substack{i\in A\\ \dist(i,\Lambda\setminus A)\leq r}}1
  &\leq \left|\{i\in A\mid \inf_{j\in\mathbb{Z}^D\setminus A}\|i-j\|_2\leq r\}\right|\nonumber\\
  &\leq\left|\{i\in A\mid \inf_{j\in\mathbb{Z}^D\setminus A}\|i-j\|_{\infty}\leq r\}\right|\nonumber\\
  &\leq 2D\left|\{(i_1,\ldots,i_D)\in \{1,\ldots,l\}^D\mid i_1\leq r\}\right|
  \leq2Dl^{D-1}r, 
\end{align}
where $\|i-j\|_{\infty}\coloneqq \max_{k=1,\ldots, D}|i_k-j_k|$.

Using $\kappa_D$ and $\kappa'_D$, we have
\begin{align}
  U_A=\sum_{i\in A,j\notin A}\|U_{ij}\|&\leq U_0\sum_{i\in A,j\notin A}(1+r_{ij})^{-(D+\delta)}
=U_0\sum_{r=1}^\infty\sum_{\substack{i\in A,j\notin A\\ r-1<r_{ij}\leq r}}(1+r_{ij})^{-(D+\delta)}\nonumber\\
&\leq U_0\sum_{r=1}^\infty r^{-(D+\delta)}\sum_{\substack{i\in A,j\notin A\\ r-1<r_{ij}\leq r}}1
\leq U_0\sum_{r=1}^\infty r^{-(D+\delta)}\kappa_Dr^{D-1}\sum_{\substack{i\in A\\ \dist(i,\Lambda\setminus A)\leq r}}1\nonumber\\
&\leq U_0\kappa_D\sum_{r=1}^\infty r^{-(1+\delta)}\min(l^D,\kappa'_Dl^{D-1}r)\nonumber\\
&\leq U_0\kappa_D\left[\kappa'_Dl^{D-1}\sum_{r=1}^l r^{-\delta}+l^D\sum_{r=l+1}^\infty r^{-(1+\delta)}\right]= U_0o(l^D). \label{suppl:residual_subsystem}
\end{align}
More precisely, it is evaluated for each $\delta$ as follows:
\begin{align}
  U_A\leq U_0O(l^D)\times\begin{cases}
    l^{-\delta} & (0<\delta<1)\\
    l^{-1}\ln l & (\delta=1)\\
    l^{-1} & (\delta>1).
  \end{cases}
\end{align}

In general, 
even if $|\Lambda\setminus\bigsqcup_{A\in\mathcal{A}}A|=o(V)$, 
the residual interaction is also subextensive because
\begin{align}
  \|U^R_{\mathcal{A}}\|
  &\leq\sum_{A\in\mathcal{A}}U_A+ \sum_{i,j\notin\bigsqcup_{A\in\mathcal{A}}A}\|U_{ij}\|\nonumber\\
  &\leq \sum_{A\in\mathcal{A}}U_A+\ab(1+\frac{\kappa_D}{\delta})U_0\left|\Lambda\setminus\bigsqcup_{A\in\mathcal{A}}A\right|=U_0o(V),
\end{align}
where we apply a similar estimation as inequality~\eqref{suppl:residual_subsystem} to derive the second inequality. 
Using this relationship, 
the following discussions can be suitably modified in cases where $\mathcal{A}$ is not an exact partition.

\subsection{Small incremental entangling (SIE) theorem}
We fix an initial state $\rho$ and a Hamiltonian $H$ with a short-range and two-body interaction, 
and write $S_A(t)$ for the (time-dependent) entropy on the subsystem $A$.
According to the Small incremental entangling (SIE) theorem~\citeSM{suppl_bravyiUpperBoundsEntangling2007,suppl_vanacoleyenEntanglementRatesArea2013}, we have~\citeSM{suppl_marienEntanglementRatesStability2016,suppl_gongEntanglementAreaLaws2017} 
\begin{equation}
  \left|\odv{S_A}{t}\right|\leq\tilde{C}\ln dU_A. \label{suppl:SIE}
\end{equation}
Here, $\tilde{C}$ is a positive constant that is independent of $d$, 
the Hamiltonian and the system size. 
We know the r.h.s.~of~\eqref{suppl:SIE} is $o(|A|)$ by~\eqref{suppl:residual_subsystem}. 

We here remark the condition of the short-range property~\eqref{suppl:short-range}. 
The authors of~\citeSM{suppl_gongEntanglementAreaLaws2017} consider a more general subsystem and require that
the interactions decay faster than $r^{-(D+1)}$ as the short-range property.
On the other hand, 
since we consider only $D$-dimensional subsystems, 
it follows from property~\eqref{suppl:bulk_property} that~\eqref{suppl:short-range} is sufficient.

\section{Properties of the canonical energy}\label{sec:suppl_canonical}
\subsection{Definitions and convexity of thermodynamic functions}
We consider a general quantum system $\mathcal{H}\cong\mathbb{C}^d$ and the Hamiltonian $H$ on it. 
We denote the Gibbs state with temperature $\beta>0$ by $\rho_H(\beta)\coloneqq e^{-\beta H}/\Tr e^{-\beta H}$. 
The corresponding free energy, internal energy, and entropy are defined as
\begin{align}
  F_H(\beta)&\coloneqq -\beta^{-1}\ln\Tr e^{-\beta H},\\
  E_H(\beta)&\coloneqq \braket<H>_{\rho_H(\beta)}=\frac{\Tr(He^{-\beta H})}{\Tr e^{-\beta H}}=\pdv{}{\beta}\beta F_H(\beta),\label{definition_of_canonical_energy}\\
  S_H(\beta)&\coloneqq \vn(\rho_H(\beta))=\beta(E_H(\beta)-F_H(\beta)). 
\end{align}
Their derivatives with respect to $\beta$ are given by
\begin{align}
  \pdv{}{\beta}F_H(\beta)&=\beta^{-2}S_H(\beta)\geq0,\label{derivative_of_free_energy}\\
  \pdv{}{\beta}E_H(\beta)&=-\braket<H^2>_{\rho_H(\beta)}+\braket<H>_{\rho_H(\beta)}^2\eqqcolon -\sigma_H^2(\beta)\leq0,\\
  \pdv{}{\beta}S_H(\beta)&=-\beta\sigma_H^2(\beta)\leq0. 
\end{align}
In particular, since $\sigma_H^2(\beta)>0$ unless $H$ is trivial, 
$\beta\in(0,\infty)$ and $E\in(E_0,\Tr H/d)$, $S\in(\ln d_0,\ln d)$ have one-to-one monotonic correspondence. 
Here, $E_0$ is the ground-state energy and $d_0$ is the degeneracy of the ground state. 
We therefore denote the internal energy $E$ and the inverse temperature $\beta$ as functions of entropy: $E_H(S)$, $\beta_H(S)$. 

These functions can also be obtained by a Legendre transformation.
The quantity $E_H(S)$ introduced in the main text is defined as
\begin{equation}
  E_{H} (S)\coloneqq \sup_{\beta>0}\left[\beta^{-1}\left(-\ln\Tr e^{-\beta H}+S\right)\right]
  =\sup_{\beta>0}\left[F_H(\beta)+\beta^{-1}S\right]. \label{Legendre_transformation_of_free_energy}
\end{equation}
A direct calculation using Eq.~\eqref{derivative_of_free_energy} shows that 
the supremum on the r.h.s.~of Eq.~\eqref{Legendre_transformation_of_free_energy} is achieved by $\beta$ such that $S=S_H(\beta)$.
Therefore, two difinitions in the main text and Eq.~\eqref{definition_of_canonical_energy} coincide if $S\in(\ln d_0,\ln d)$.
We note that $E_{H} (S)$ is equal to $E_0$ if $S\leq\ln d_0$.
Moreover, the definition of $\beta^{-1}$ in the main text is also consistent with the definition of the same symbol introduced here.

$E_H$ is a convex function of $S$ because
\begin{equation}
  \pdv[order=2]{E_H}{S}=\pdv{\beta^{-1}}{S}=\frac{1}{\beta^3\sigma_H^2}>0.
\end{equation}
Also, the free energy is continuous with respect to the Hamiltonian (see, e.g.,~\citeSM{suppl_ruelleStatisticalMechanicsRigorous1999}): 
\begin{equation}
  |F_H(\beta)-F_{H'}(\beta)|\leq\|H-H'\|\ (\forall\beta)\label{suppl:free_energy}. 
\end{equation}

\subsection{Maximum entropy principle and minimum energy principle}
The maximum entropy principle~\citeSM{suppl_jaynesInformationTheoryStatistical1957} states that for a given energy expectation value, 
the state that maximizes entropy is a Gibbs state. 
In the positive temperature regime, 
this property is equivalent to the minimum energy principle, 
which states that the state that minimizes the energy expectation value for a fixed entropy is the Gibbs state. 
We adopt this principle as inequality~\eqref{inequalities_maxent} in the main text, which is represented as
\begin{equation}
  E_H(\vn(\sigma))\leq\braket<H>_\sigma.\label{suppl:minimum_energy}
\end{equation}

These properties follow from the non-negativity of the Kullback--Leibler divergence $D(\sigma\|\rho)$. 
In fact, we have 
\begin{equation}
  0\leq D(\sigma\|\rho_H(\beta))=-\vn(\sigma)+\beta\braket<H>_\sigma+\Tr e^{-\beta H}
  =S_H(\beta)-\vn(\sigma)+\beta(\braket<H>_\sigma-E(\beta)). \label{suppl:div_from_gibbs}
\end{equation}
We get the maximum entropy principle by taking $\beta$ as $E(\beta)=\braket<H>_\sigma$, 
and get the minimum energy principle by dividing $\beta>0$ and minimizing the rightmost side. 
The equality of~\eqref{suppl:div_from_gibbs} is met if and only if $\sigma=\rho_H(\beta)$. 

\subsection{Conditions for local equilibrium ensembles}
We justify the two assumptions for the local equilibrium ensemble: 
\begin{align}
  &V\epsilon\simeq\braket<H>_{\rho^\lEq}\simeq \sum_AE_{H_A} (\vn(\rho^\lEq_A)), \label{suppl:assumption_ensemble}\\
  &(\beta_A^\lEq)^{-1}\coloneqq \beta_A^{-1}(\vn(\rho^\lEq_A))\leq\beta_0^{-1}\ (\forall A\in\mathcal{A}), \label{suppl:assumption_subsystems}
\end{align}
which are used in the main text.  
We note that 
\begin{equation}
  \braket<H>_{\rho^\lEq}\simeq\sum_A\braket<H_A>_{\rho^\lEq}\geq\sum_AE_{H_A} (\vn(\rho^\lEq_A))\label{suppl:lower_bound}
\end{equation}
holds because the residual interaction is subextensive and inequality~\eqref{suppl:minimum_energy} holds. 

\subsubsection{The product of local Gibbs state}
If $\rho^\lEq=\otimes_A\rho_{H_A}(\beta_A)$, then the equality in the right inequality of~\eqref{suppl:lower_bound} holds. 
Therefore, the first assumption~\eqref{suppl:assumption_ensemble} is satisfied by $(\beta_A)_A$
as the energy expectation is almost $V\epsilon$. 
The second assumption~\eqref{suppl:assumption_subsystems} is equivalent to the condition $\beta_A^{-1}\leq\beta_0^{-1}$. 
If $\epsilon$ corresponds to a positive temperature, we can take such $\beta_0$.

\subsubsection{The canonical state}
Consider the case $\rho^\lEq=\rho_{H}(\beta)$. 
Here, $\beta$ is taken to satisfy $E_H(\beta)\simeq V\epsilon$. 
First, we prove that the first assumption~\eqref{suppl:assumption_ensemble} is valid. 
By inequality~\eqref{suppl:lower_bound}, 
it is sufficient to show the inverse inequality of~\eqref{suppl:lower_bound}. 
By the definition of the canonical energy, we have
\begin{align}
  \sum_AE_{H_A} (\vn(\rho_{H}(\beta)_A))
  &=\sum_A\sup_{\beta_A>0}\left[\beta_A^{-1}\left(-\ln\Tr e^{-\beta_A H_A}+\vn(\rho_{H}(\beta)_A)\right)\right]\nonumber\\
  &\geq\sup_{\beta_*>0}\sum_A\left[\beta_*^{-1}\left(-\ln\Tr e^{-\beta_* H_A}+\vn(\rho_{H}(\beta)_A)\right)\right]\nonumber\\
  &=\sup_{\beta_*>0}\beta_*^{-1}\sum_A\left[-\ln\Tr e^{-\beta_* H_A}+\vn(\rho_{H}(\beta)_A)\right]\nonumber\\
  &=\sup_{\beta_*>0}\beta_*^{-1}\left(-\ln\Tr e^{-\beta_* \sum_AH_A}+\sum_A\vn(\rho_{H}(\beta)_A)\right)\nonumber\\
  &=\sup_{\beta_*>0}F_{\sum_AH_A}(\beta_*)+\beta_*^{-1}\sum_A\vn(\rho_{H}(\beta)_A). \label{suppl:canonical_subsystem}
\end{align}
It follows from the continuity of free energy~\eqref{suppl:free_energy}:
\begin{equation}
 |F_{\sum_AH_A}(\beta_*)-F_{H}(\beta_*)|\leq\|U^R_{\mathcal{A}}\|\ (\forall\beta_*)\label{suppl:residual_energy},
\end{equation}
and the subadditivity of entropy:$\sum_A\vn(\rho_{H}(\beta)_A)\geq \vn(\rho_{H}(\beta))$ that
\begin{align}
  \mbox{(r.h.s.~of~\eqref{suppl:canonical_subsystem})}&\geq\sup_{\beta_*>0}F_{H}(\beta_*)+\beta_*^{-1}\vn(\rho_{H}(\beta))+\|U^R_{\mathcal{A}}\|\nonumber\\
  &\simeq\sup_{\beta_*>0}F_{H}(\beta_*)+\beta_*^{-1}\vn(\rho_{H}(\beta))=E_H(\vn(\rho_{H}(\beta)))=\braket<H>_{\rho_{H}(\beta)}. 
\end{align}

For a translationally invariant system, 
$\braket<H_A>$ is independent of $A$, 
so that the left-hand side of~\eqref{suppl:assumption_subsystems} is independent of $A$. 
In this case, inequality~\eqref{suppl:assumption_subsystems} holds if $\beta_0^{-1}$ is sufficiently larger than $\beta^{-1}$.

\subsubsection{The microcanonical state}
We take the microcanonical state $\rho^\mc$ of the energy shell around $\epsilon$ for $\rho^\lEq$. 
From the ensemble equivalence of thermodynamic functions~\citeSM{suppl_ruelleStatisticalMechanicsRigorous1999,suppl_tasaki_local_2018}, 
the same argument holds as for the canonical states. 
More specifically, by an argument similar to that in the previous section,
we have
\begin{equation}
  V\epsilon\simeq\braket<H>_{\rho^\mc}\gtrsim\sum_AE_{H_A} (\vn(\rho^\mc_A))
  \gtrsim E_H(\vn(\rho^\mc)). \label{suppl:can_mc}
\end{equation}
Therefore, we only have to show the inverse direction of this inequality. 

It is known~\citeSM{suppl_ruelleStatisticalMechanicsRigorous1999,suppl_tasaki_local_2018} 
that there exist thermodynamic limits of free energy and entropy: 
\begin{align}
  \lim_{V\to\infty}\frac{1}{V}F_H(\beta)&\eqqcolon f(\beta),\\
  \lim_{V\to\infty}\frac{1}{V}\vn(\rho^\mc)&\eqqcolon s(\epsilon),
\end{align}
which are continuous, convex, 
and connected via the Legendre transformation: 
\begin{equation}
  f(\beta)=\beta^{-1}\inf_{u}(\beta u-s(u)). 
\end{equation}
If $\epsilon$ corresponds to a positive temperature, 
we can take $\beta(\epsilon)>0$ satisfying
\begin{equation}
  f(\beta(\epsilon))=\beta(\epsilon)^{-1}(\beta(\epsilon)\epsilon-s(\epsilon))
\end{equation}

Using this, we can evaluate the rightmost side of~\eqref{suppl:can_mc} as
\begin{align}
  \liminf_{V\to\infty}\frac{1}{V}E_H(\vn(\rho^\mc))
  &\geq\sup_{\beta>0}\liminf_{V\to\infty}\frac{1}{V}\ab(F_{H}(\beta)+\beta^{-1}\vn(\rho^\mc))\nonumber\\
  &=\sup_{\beta>0}f(\beta)+\beta^{-1}s(\epsilon)
  \geq f(\beta(\epsilon))+\beta(\epsilon)^{-1}s(\epsilon)=\epsilon. 
\end{align}
Therefore, assumption~\eqref{suppl:assumption_ensemble} holds for the microcanonical state. 
Assumption~\eqref{suppl:assumption_subsystems} can be treated in the same way as for the canonical state.

\section{Applicability to bosonic and fermionic systems}\label{sec:suppl_FBsystem}
In this section, 
we discuss in detail the extension of our results in spin systems to fermionic and bosonic systems. 
Let us decompose the $1$-particle state space $\mathcal{K}$ into modes $\lambda\in\Lambda'$, 
i.e., $\mathcal{K}=\oplus_{\lambda\in\Lambda'}\mathcal{K}_{\lambda}$.
We denote the Fock space of fermions and that of bosons as $\mathcal{F}_f(\mathcal{K})$ and $\mathcal{F}_b(\mathcal{K})$, respectively.

We assume that each mode $\lambda\in\Lambda'$ is a pair of sites $i\in\Lambda$ on a hypercubic lattice 
and internal degrees of freedom $\sigma\in\Lambda_{\mathrm{int}}$ at each site. 
We divide the lattice into small (but diverging in the thermodynamic limit) hypercubes: $\Lambda=\bigsqcup_{A\in\mathcal{A}}A$, as in spin systems,  
and define an on-site state space $\mathcal{K}_i$ and a subsystem state space $\mathcal{K}_A$ as
\begin{align}
  \mathcal{K}_i&:=\oplus_{\sigma\in\Lambda_{\mathrm{int}}}\mathcal{K}_{(i,\sigma)}\\
  \mathcal{K}_A&:=\oplus_{i\in A} \mathcal{K}_i.
\end{align}

Considering only an on-site interaction or an interaction involving two sites,
the Hamiltonian takes the same form as in the main text: $H=\sum_{i\in\Lambda}h_i+\sum_{i\neq j\in\Lambda}U_{ij}$. 
Here $h_i$ is the on-site Hamiltonian at site $i$, 
and $U_{ij}$ is the interaction between distinct sites $i$ and $j$. 
The Hamiltonian of each subsystem and the residual interaction are also defined in a similar manner. 
We note that if a particle number is conserved, 
we should add those quantities to the variables in the statistical ensemble. 
This can be done if the total state space $\mathcal{H}$ is chosen to be a subspace of $\mathcal{F}_{\bullet}(\mathcal{K})$. 

\subsection{Fermionic systems}
In fermionic systems, there is no canonical isomorphism between $\mathcal{F}_f(\mathcal{K})$ and $\otimes_{i\in\Lambda}\mathcal{F}_f(\mathcal{K}_i)$ due to anticommutativity. 
Instead, we take a unitary isomorphism $U_A:\mathcal{F}_f(\mathcal{K})\cong\mathcal{F}_f(\mathcal{K}_A)\otimes\mathcal{F}_f(\mathcal{K}_{\Lambda\setminus A})$ 
so that creation (annihilation) operator $c_i^{\dag}(f)$ $(c_i(f))$ on $i\in A$ is mapped to $c_i^{\dag}(f)\otimes I$ $(c_i(f)\otimes I)$. 
Through this isomorphism, the same argument can be made if the reduced state is defined as: 
\begin{equation}
  \rho_A:=\Tr_{\Lambda\setminus A}U_A\rho U_A^\dag.
\end{equation}
Therefore, Theorem~\ref{thm:universal_bound} is applicable to the following class of Hamiltonians, including the Fermi--Hubbard model: 
\begin{equation}
  H=\sum_{ij,\sigma}t_{ij}c_{i\sigma}^\dag c_{j\sigma}+\sum_{ij}U_{ij}n_in_j+\sum_{ij}U'_{ij}\bm{S}_i\cdot\bm{S}_j, 
\end{equation}
where $\sigma$ represents the spin of particles, $n_i=\sum_\sigma c_{i\sigma}^\dag c_{i\sigma}$ is the number operator and $\bm{S}_i$ is the second-quantized spin operator.
Here, $t,U,$ and $U'$ are assumed to decay similarly as in spin systems (see Eq.~\eqref{short-range} in the main text). 
Since the SIE theorem is known to hold for this case as well~\citeSM{suppl_marienEntanglementRatesStability2016}, 
Corollary~\ref{cor:MITE_means_passive} also holds.

\subsection{Bosonic systems}
In bosonic systems, there exists a canonical isomorphism $\mathcal{F}_b(\mathcal{K})\cong\otimes_{i\in\Lambda}\mathcal{F}_b(\mathcal{K}_i)$. 
However, the dimension of the subsystem depends on the total particle number $N$ due to the lack of restrictions on the local number of particles, 
which causes problems that are absent in spin systems. 
In particular, the norm of the residual interaction $U^R_{\mathcal{A}}$ does not necessarily become subextensive due to the effect of localized states at the boundary.

A similar argument holds if one imposes the additional assumption that the operation does not cause Bose-Einstein condensation. 
For example, 
for a hard-core boson system, 
which is equivalent to a spin-1/2 system, 
the main theorem and the corollary are also applicable.

In general, Theorem~\ref{thm:universal_bound} is valid 
if the assumptions~\eqref{suppl:assumption_ensemble},~\eqref{suppl:assumption_subsystems} are justified 
and $\braket<U^R_{\mathcal{A}}>_{f(\rho)}\simeq0$ holds. 
In the discussion in the previous section, the key relations are 
the first equality in~\eqref{suppl:lower_bound} and~\eqref{suppl:residual_energy}. 
Physically, these relations imply the additivity of the (free) energy, 
which is satisfied in the usual models in statistical mechanics.
On the other hand, 
the validity of $\braket<U^R_{\mathcal{A}}>_{f(\rho)}\simeq0$ and 
the applicability of Corollary~\ref{cor:MITE_means_passive} are more nontrivial. 

For example, 
we consider the Bose--Hubbard Hamiltonian:
\begin{equation}
  H=-t\sum_{\langle ij\rangle}(b_i^\dag b_j+b_ib_j^\dag)+\frac{U}{2}\sum_in_i(n_i-1)-\mu\sum_in_i, 
\end{equation}
where $\langle ij\rangle$ represents a nearest neighbor pair 
and $n_i=b_i^\dag b_i$ is the number operator. 
According to Eq.~(6) in Ref.~\citeSM{suppl_kuwaharaEffectiveLightCone2024}, 
the finite-time evolution of low-density states in such a system can be approximated with error $V^{-a}$
by a Hamiltonian restricted to a space 
where the local particle density is not more than $q=O(\textrm{polylog}(V))$. 

For such a truncated Hamiltonian, 
we can estimate the norm of the residual interaction as 
\begin{align}
  \|U_A\|&\leq|t|\sum_{\langle\langle ij\rangle\rangle}\|b_i^\dag b_j+b_ib_j^\dag\|\\
  &\leq|t|\sum_{\langle\langle ij\rangle\rangle}\|n_i\|+\|n_j\|\\
  &=O(1)Vql^{-1}=VO(\textrm{polylog}(V))l^{-1},  
\end{align}
where $\langle\langle ij\rangle\rangle$ represents a nearest neighbor pair 
between $A$ and a different subsystem. 
Therefore, if the size of subsystem $l^D$ is taken to be the power of $V$, 
we can neglect the decrease in entropy of subsystem by inequality~\eqref{suppl:SIE}, 
and therefore Corollary~\ref{cor:MITE_means_passive} holds.

\bibliographystyleSM{apsrev4-2}
\bibliographySM{ergotropySM}

\end{document}